\documentclass[aps,prl,twocolumn,longbibliography,superscriptaddress]{revtex4-1}

\usepackage[pdftex]{hyperref,graphicx}
\usepackage{amssymb}
\usepackage{amsmath}
\usepackage{xcolor}
\usepackage[caption=false]{subfig}

\hypersetup{
    pdftitle = {Detecting topological invariants in chiral symmetric insulators via losses},
    pdfauthor = {Tibor Rakovszky, Janos K. Asboth, Andrea Alberti},
    colorlinks=true,linkcolor=blue,citecolor=blue,filecolor=blue,urlcolor=blue
  }

\newcommand{\expect}[1]{\langle{#1}\rangle}
\newcommand{\ket}[1]{|{#1}\rangle}
\newcommand{\bra}[1]{\langle{#1}|}
\newcommand{\braket}[2]{\langle{#1}|{#2}\rangle}
\newcommand{\abs}[1]{\left|{#1}\right|}
\newcommand{\tr}{\textrm{Tr }}
\newcommand{\dblavg}[1]{\langle\kern-1.5pt\langle{#1}\rangle\kern-1.5pt\rangle}

\newcommand{\Hflat}{\hat{Q}}
\newcommand{\XX}{\hat{X}}
\newcommand{\Qp}{\hat{Q}_+}
\newcommand{\Qm}{\hat{Q}_-}

\newcommand{\UU}{\hat{U}}

\newcommand{\HH}{\hat{H}}
\newcommand{\Heff}{\hat{H}_\text{eff}}

\newcommand{\GG}{\hat{\Gamma}}
\newcommand{\MM}{\hat{M}}

\newcommand{\PA}{\hat{P}_A}
\newcommand{\PB}{\hat{P}_B}

\newcommand{\encapsulateMath}[1]{\raisebox{0pt}[0pt][0pt]{#1}}

\renewcommand\textemdash{\leavevmode\unskip\kern0.8pt\rule[0.19\baselineskip]{8pt}{0.4pt}\kern1pt\ignorespaces}

\makeatletter
\newcommand{\longdash}[1]{{\emph{#1}}.\textemdash}
\renewcommand\section{\@startsection {section}{1}{9.24774pt}%
                                   {1pt}%
                                   {0em}%
                                   {\raggedright\longdash}}
\DeclareRobustCommand{\@seccntformat}[1]{%
  \def\temp@@a{#1}%
  \def\temp@@b{section}%
  \ifx\temp@@a\temp@@b
  \csname the#1\endcsname .\quad%
  \else
  \csname the#1\endcsname\quad%
  \fi
}
\makeatother

\begin{document}

\title{Detecting topological invariants in chiral symmetric
 insulators via losses}
\author{Tibor Rakovszky}
\affiliation{Max-Planck-Institut f{\"u}r Physik komplexer Systeme,
  N{\"o}thnitzer Str. 38, 01187 Dresden, Germany}
\author{J\'anos K. Asb\'oth}
\affiliation{Institute for Solid State
  Physics and Optics, Wigner Research Centre for Physics, 
Hungarian Academy of
  Sciences, H-1525 Budapest P.O. Box 49, Hungary}
\author{Andrea Alberti} 
\affiliation{Institut f{\"u}r Angewandte Physik,
  Universit{\"a}t Bonn, Wegelerstr. 8, D-53115 Bonn, Germany} 
\date{December 2016}
\begin{abstract}
We show that the bulk winding number characterizing one-dimensional topological insulators with chiral symmetry can be detected from the displacement of a single particle, observed via losses. 
Losses represent the effect of repeated weak measurements on one sublattice only, which interrupt the dynamics periodically.
When these do not detect the particle, they realize negative measurements.
Our repeated measurement scheme covers both time-independent and periodically driven (Floquet) topological insulators, with or without spatial disorder.
In the limit of rapidly repeated, vanishingly weak measurements, our scheme  
describes non-Hermitian Hamiltonians, as the lossy 
Su-Schrieffer-Heeger model of Phys. Rev. Lett. 102, 065703 (2009).
We find, contrary to intuition, that the time needed to
detect the winding number can be made shorter by
decreasing the efficiency of the measurement. 
We illustrate our results on a discrete-time quantum
walk, and propose ways of testing them experimentally.
\end{abstract}
\maketitle

Topological insulators~\cite{hasan2010colloquium} are materials whose
bulk is gapped, and is characterized by a topological invariant.
Depending on the dimensionality of the system and the
discrete symmetries it possesses, this invariant can be a Chern
number, a winding number, or some other mathematical
index~\cite{budich2013adiabatic,schnyder_2016}.
The bulk invariant predicts a number of robust low-energy eigenstates at the edges via the so-called bulk-boundary
correspondence~\cite{ryu_tenfold2010,teo_kane2010}.
In one dimension, these are bound states at the
ends of the topological insulator wire.
The energy of these states is protected against perturbations
due to either particle-hole symmetry, as for the Majorana
fermions~\cite{mourik2012signatures}
which might be used to store qubits, or to chiral (sublattice) symmetry, as for bound states at domain walls in polyactylene molecules~\cite{heeger_nobel}.
Hence, bulk topological invariants control the robust properties
of topological insulators.

Of special interest are experiments implementing topological insulators 
with artificial matter setups, where
bulk topological invariants can not only be inferred from the presence of edge states, but also measured directly~\cite{price_prb_2016}.
Recently, such experiments have been performed using cold atoms in
optical lattices~\cite{jotzu2014experimental,
  aidelsburger2015measuring, flaschner2016experimental,li2016bloch,weitz_2016_experimental,Meier:2016},
and using light~\cite{hafezi2013imaging,zeuner2015observation,Bleckmann:2016} or
microwaves~\cite{poli:2015} in photonic crystal-like structures.
These setups are ideal model systems for
topological insulators, often employing periodic driving as a tool to
engineer the effective Hamiltonian.
Topological invariants are detected by measuring the displacement of a
cloud of
particles~\cite{aidelsburger2015measuring,jotzu2014experimental}, or by
interferometric schemes~\cite{atala2013direct}.
Alternatively, the topological invariant can be observed 
by attaching leads to the system, and measuring the reflection
amplitudes for scattering off the bulk~\cite{fulga2012scattering,tarasinski2014scattering,
  fulga2016scattering,hu2015measurement}.
This last approach has recently been applied to detect winding numbers
in a one-dimensional quantum walk, an ideal system for periodivally driven topological
insulators~\cite{barkhofen2016measuring}.

Topological invariants can also appear in non-Hermitian systems, as
predicted by Rudner and Levitov~\cite{rudner2009topological}, and
recently realized experimentally~\cite{zeuner2015observation}.
In that scheme, the Su-Schrieffer-Heeger
(SSH)~\cite{heeger_nobel,asboth2016book} model \textemdash a nearest-neighbor-hopping Hamiltonian with topological invariants due to chiral symmetry \textemdash is modified
by adding losses to every second site.
The average distance traveled by a particle, initialized on a nonlossy site, before it is lost, is an integer coinciding with the winding number of the original SSH model.
However, whether a similar correspondence holds
generally, for any chiral symmetric system in one dimension, has so far remained an open question.

In this Letter, we show that the expected displacement of a single particle, measured through losses, is given by the bulk topological invariant for any chiral symmetric one-dimensional topological insulator, even in the presence of periodic driving, with or without disorder.
Our approach is formulated in the language of periodically driven systems, but by including weak measurements we are also able to recover the
case of time-independent multi-band Hamiltonians.
Note that we use losses to detect topological invariants of the unitary dynamics \textemdash unlike other work where topological invariants are engineered through dissipation~\cite{bardyn2013topology,budich2015dissipative}.

\section{The setup} We consider a generic one-dimensional lattice system of noninteracting particles. 
To describe the state of a single particle we use position eigenstates $\ket{x,c}$, where $x=1,\ldots,L$ denote the unit cells and $c=1,\ldots,2N$ are the states forming a basis of a single unit cell. 
These can be $2N$ different sites, but can also be regarded as $2N$ internal states of the particle~\footnote{If the number of internal states involved in the system's dynamics is odd, it would not be possible to have a system that is both chiral symmetric and gapped.}.

The dynamics is given by a periodically driven Hamiltonian $\HH(t) = \HH(t+T)$. 
This trivially includes time independent systems, where $T$ can be chosen arbitrarily. 
The net time evolution during one driving period is described by the unitary operator $\UU = \mathbb{T} e^{-i\int_0^T \mathrm{d}t \HH(t)}$, where $\mathbb{T}$ denotes time ordering.

It is often useful to think of the dynamics in terms of a time-independent \emph{effective Hamiltonian} $\Heff$, defined by
\begin{equation}
\label{eq:time_ordered}
\UU = e^{-i\Heff T}.
\end{equation}
The eigenvalues of $\Heff$, called \emph{quasienergies} and denoted $E_n$, are periodic with period $2\pi/T$ and can be chosen to lie in the interval $E_n\in[-\pi/T,\pi/T)$. 
The corresponding eigenstates are stationary states of the discrete time evolution that only acquire a phase $e^{-iE_n T}$ during a full cycle.
Note that for time-independent systems, the effective Hamiltonian and the usual Hamiltonian coincide. 
In the following we use dimensionless energy $\varepsilon_n = E_nT$.

\section{Chiral symmetry and winding number} To enable the system to have nontrivial topological phases, we need to impose some constraint on it, which in
this work is \emph{chiral symmetry}.
Chiral symmetry for lattice Hamiltonians is also known as sublattice
symmetry, since it can be defined by first grouping all internal
states into two sets, the \emph{sublattices} $A$ and $B$. We define these
by projectors
\begin{align}\label{eq:def_sublattices}
  \PA &=  \sum_{x\in\mathbb{Z}} \sum_{a=1}^{N} \ket{x,a}\bra{x,a},&
\PB &= \hat{1} - \PA,
\end{align}
where $\hat{1}$ is the identity.  
Chiral symmetry means that the effective Hamiltonian has no matrix
elements between states on the same sublattice, i.e.
\begin{align}\label{eq:cs_def}
\GG\Heff\GG &= -\Heff& \text{for }\GG &= \PA - \PB.
\end{align}
The chiral symmetry operator $\GG$, defined above, acts on each unit cell separately and satisfies $\GG^{-1} = \GG^\dagger = \GG$. In fact we could relax the above condition and replace it with the less strict requirement $\GG\UU\GG = \UU^\dagger$, which is implied by Eq.~\eqref{eq:cs_def}.

An immediate consequence of chiral symmetry is that the eigenstates
come in pairs $\{\ket{n},\GG\ket{n}\}$, with quasienergies
$\{-\varepsilon_n,\varepsilon_n\}$.
We will use the projectors onto the upper and lower half of the
spectrum,
\begin{align}\label{eq:eigenstate_projectors}
\Qm &= \sum_{-\pi \leq \varepsilon_n \leq 0} \ket{n}\bra{n},&\Qp &= \GG\Qm\GG.
\end{align}

Chiral symmetry allows the system to have nontrivial bulk topological phases. These are characterized by an integer winding number, defined in its most general form~\cite{mondragon2014topological} as
\begin{equation}\label{eq:realspace_winding_def}
\nu = \frac{1}{L} \tr \left\{ \PB\Hflat\PA\,[\XX,\PA\Hflat\PB]\right\},
\end{equation}
where $\XX = \sum_{x\in\mathbb{Z}}\sum_{c=1}^{2N} x
\ket{x,c}\bra{x,c}$ is the position operator, and $\Hflat = \Qp - \Qm$
is the flat-band limit of the (effective) Hamiltonian, defined in terms of
the projectors in Eq.~\eqref{eq:eigenstate_projectors}.
In the presence of translational invariance, the above formula for the winding number
reduces to its usual definition in quasimomentum space.
However, the real-space formula for $\nu$ is also valid for disordered
systems. Physically, it measures the difference of the electric
polarisations of the two sublattices~\cite{mondragon2014topological}.

\section{Weak partial measurement after each period} 
\begin{figure}[h!]
 \centering
  \includegraphics[width=0.4\textwidth]{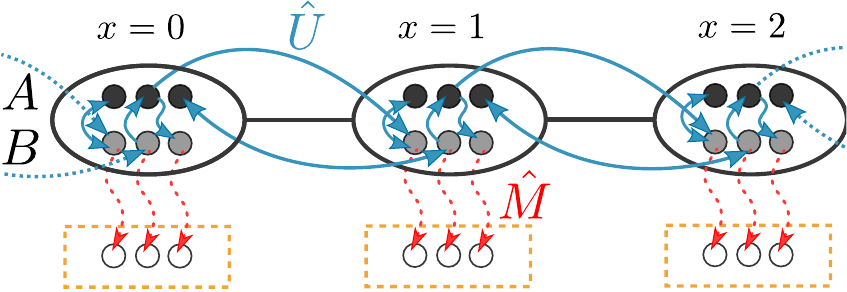}
 \caption{Weak measurement of position on one sublattice only, following each unitary step.
First, sites of sublattice B (light grey circles) are coupled (red dashed lines) for a fixed time to ancillary sites.
Then, the population of each ancillary site is measured.
If all ancillary sites are found empty, we have a negative measurement, and the next unitary step follows.
The blue arrows represent matrix elements of the effective Hamiltonian $\Heff$.}
  \label{fig:measurement}
 \end{figure}

To detect the winding
number, we initalize a single particle on a site on sublattice $A$,
then apply the unitary $\UU$ repeatedly, with each application followed by
a partial position measurement.
We call this a \emph{partial measurement}, because it measures position only on
sublattice $B$, while avoiding any interaction with the sites of sublattice $A$.
The measurement operation is parametrized by its efficiency $0<p_M\leq1$.
\emph{Weak measurements} ($p_M < 1$) can be realized by coupling the sites of sublattice $B$ to initially unoccupied ancillary sites for a fixed, short time, and then measuring the occupation of the ancillary sites, as shown in Fig.~\ref{fig:measurement}.
The measurement can yield a positive result, with conditional probability $p_M$, in which case we detect the position $x$ of the particle and halt observing its quantum evolution.
If the measurement returns with a negative result \textemdash \emph{a negative measurement} \textemdash then we continue with the next unitary driving cycle, followed by the next measurement phase, and repeat this procedure until a successful detection occurs. 

A practical tool to calculate relevant quantities with such
repeat-until-detection quantum dynamics is the
\emph{conditional wavefunction}.
To define it, we first introduce the linear but not unitary operator
$\MM$, representing the effect of the negative measurement on the
wavefunction, as
\begin{align}
  \MM &= \PA + \sqrt{1-p_M} \PB.
\end{align}
The conditional wavefunction of the system
after $j$ driving cycles but before
the $j$-th measurement operation is
\begin{align}
\ket{\tilde\Psi(t=jT)} &= [\UU\MM]^{j-1} \UU \ket{\Psi(0)} \quad
\text{for } j \in \mathbb{N}.
\label{eq:conditional_time_evolution}
\end{align}
The norm $\braket{\tilde\Psi(jT)}{\tilde\Psi(jT)}$ is the probability
that the particle was not detected during the first $j-1$ measurements.

By allowing for weak measurements, with $p_M < 1$,
we can also cover the case of nondriven, time-independent systems, as
in Ref.~\onlinecite{rudner2009topological}.
There, the chiral symmetric Hamiltonian $\HH$ is modified by an
imaginary term describing losses,
\begin{align}
\label{eq:nondriven_def}
  \HH &\to \HH - i\frac{\gamma}{2} \PB,
\end{align}
where $\gamma$ is a decay rate.
Trotterization of the corresponding time evolution is equivalent to 
Eq.~\eqref{eq:conditional_time_evolution}  in the limit $T\to 0$, with
$p_M = \gamma T$ and $\UU = e^{-i\HH T}$.

\section{Winding number from average displacement} We are interested in the displacement of the particle and the \emph{dwell time} \textemdash the time it spends in the system before it is measured -\textemdash averaged over many repetitions of the measurement process with the same initial state $\ket{x,a}$. The probability of measuring it in $\ket{y,b}$ after $j$ steps is
\begin{equation}
s_{(x,a)\to(y,b)}(j) = p_M \abs{\bra{y,b} [\UU \MM]^{j-1}\UU\ket{x,a}}^2.
\end{equation}
We define the average displacement and dwell time as
\begin{align}\label{eq:single_avgs1}
  \expect{\Delta x}_{(x,a)} &\equiv \sum_{j\in\mathbb{Z}^+}
\sum_{y=1}^L (y-x) \!\!
\sum_{b=N+1}^{2N} s_{(x,a)\to(y,b)}(j),
 \\
\label{eq:single_avgs2}  \expect{t}_{(x,a)} &\equiv T \sum_{j\in\mathbb{Z}^+}
\sum_{y=1}^L j \!\!
\sum_{b=N+1}^{2N}s_{(x,a)\to(y,b)}(j).
\end{align}

To get a general result, valid for arbitrary $N$ and for spatial disorder, we also need to average over all states on sublattice $A$ where the particle is initially prepared. We define these double-averaged quantities as
\begin{align}\label{eq:double_avgs}
\dblavg{\Delta x} = \frac{\sum_{x,a} \expect{\Delta x}_{(x,a)}}{NL}, \;\quad
\dblavg{t} = \frac{\sum_{x,a} \expect{t}_{(x,a)}}{NL}.
\end{align}
Note that for a large, disordered sample, averaging over all initial sites is expected to give the same result as averaging over different disorder realizations.
Note also that for translationally invariant systems $\expect{\Delta x}_{(x,a)}$ and $\expect{t}_{(x,a)}$ are independent of the initial position, and in this case the averaging over $x$ can be omitted. 

To analytically compute both $\dblavg{\Delta x}$ and $\dblavg{t}$, we
write the conditional wavefunction as
\begin{equation}\label{eq:expand_time_evolved_state}
\UU \left[\MM\UU\right]^{j-1}\ket{x,a} = \sum_n \left[ \alpha_n^{(x,a)}(j) \ket{A}_n + \beta_n^{(x,a)}(j) \ket{B}_n \right],
\end{equation}
where the states $\ket{A}_n = (\ket{n} + \GG\ket{n}) / \sqrt{2}$ and $\ket{B}_n = (\ket{n} - \GG\ket{n}) / \sqrt{2}$ have support on sublattices $A$ and $B$ respectively, and the sum is taken over the lower half of the spectrum. The coefficients evolve in time as
\begin{subequations}\label{eq:n_sector_time_ev}
\begin{align}
\alpha_n(j\!+\!1) &= \alpha_n(j)\cos\varepsilon_n-i\beta_n(j)\sqrt{1-p_M}\sin\varepsilon_n  \\
\beta_n(j\!+\!1) &= \beta_n(j)\sqrt{1-p_M}\cos\varepsilon_n-i\alpha_n(j)\sin\varepsilon_n
\end{align}
\end{subequations}
with indices $(x,a)$ omitted, since these only appear in the initial
condition $\alpha_n^{(x,a)}(0) = \sqrt{2}\,\braket{n}{x,a}$. Note that
$\beta_n^{(x,a)}(0) = 0$ since the initial state has support on
sublattice $A$. We read off from
Eq.~\eqref{eq:n_sector_time_ev} that for modes at quasienergies
$\varepsilon_n=0$ or $\varepsilon_n=\pi$ the coefficient $\alpha^{(x,a)}_n(j)$ remains constant in time. Therefore, these
are dark states of the lossy dynamics and if the particle has
some initial overlap with them, then it has a finite probability of
staying in the system forever.

To compute the averages defined in
Eqs.~\eqref{eq:single_avgs1} and~\eqref{eq:single_avgs2} we need the
coefficients $\beta_n^{(x,a)}(j)$, which can be obtained by solving
Eq.~\eqref{eq:n_sector_time_ev} as we show in the Supplemental
Material. Substituting the result into Eqs.~\eqref{eq:double_avgs} and
summing over the discrete time $j$ (which can be done as long as there are no dark
states in the spectrum) we obtain a compact formula for
the double-averaged displacement (derivation in the Supplemental
Material):
\begin{equation}\label{eq:deltax_finalform}
\dblavg{\Delta x} = \frac{2}{NL} \tr \left\{\XX\GG\Qm\right\}.
\end{equation}
Note that $\XX\GG = \PA\XX\PA - \PB\XX\PB$ is the difference of the
projections of the position operator \textemdash i.e.~electric polarisation \textemdash onto the two sublattices. Thus,
the above formula is clearly related to the sublattice polarisation of
Eq.~\eqref{eq:realspace_winding_def}. Indeed, after some further
algebraic manipulations we find
\begin{equation}\label{eq:mainstatement}
\dblavg{\Delta x} = \nu / N.
\end{equation}
This is our main result that relates the average displacement of a
single particle in the lossy system to the winding number associated with the
unitary time evolution.
It is valid in the same form for static systems,
where position is measured via losses as per
Eq.~\eqref{eq:nondriven_def}.
Note that to apply either Eq.~\eqref{eq:realspace_winding_def} or
Eq.~\eqref{eq:deltax_finalform} to a finite system with
periodic boundary conditions, one has to use an appropriately modified
definition of the position
operator~\cite{RestaPosition,Prodan_noncommutative}.

\section{Dwell time and quantum Zeno effect} We also find compact formulas
for the average dwell time using Eqs.~\eqref{eq:expand_time_evolved_state} and~\eqref{eq:n_sector_time_ev}. We detail the derivation in the Supplemental Material, and just discuss the results here. First, for strong measurements, $p_M=1$, the
average dwell time can be expressed, in the thermodynamic limit of
$L\to \infty$, using the density of states $\rho(\varepsilon)$ as
\begin{equation}
\label{eq:decay_time_pm=1}
\dblavg{t}{\Big{|}}_{p_M=1} = \frac{T}{N} \int_{\varepsilon=0}^{\pi} \frac{\rho(\varepsilon)}{\sin^2\varepsilon}
\mathrm{d}\varepsilon  \equiv T \tau,
\end{equation}
where we introduced the shorthand $\tau$ for the integral. 
For weak measurements, this result is modified as
\begin{equation}
\label{eq:dwell_time_result}
  \dblavg{t} = T\left[\frac{p_M }{(1+\sqrt{1 - p_M})^2}\, \tau  
  + \frac{2\sqrt{1-p_M}}{p_M}\right]. 
\end{equation}

The average dwell time can become long, or even diverge, in the
presence of almost-dark states: in Eq.~\eqref{eq:decay_time_pm=1} the
integral is dominated by states near $\varepsilon\approx 0$ and
$\varepsilon\approx\pi$. These states can occur not only near the
topological phase transition but also due to strong disorder.  For
these states the transition amplitude from sublattice $A$ to $B$
during a single step is infinitesimal. As a consequence, repeatedly
measuring the particle's presence on sublattice $B$ can prevent it
from ever occupying it, similarly to the well-known quantum Zeno
effect.

\begin{figure}[h!]
 \centering
 \includegraphics[width=0.45\textwidth]{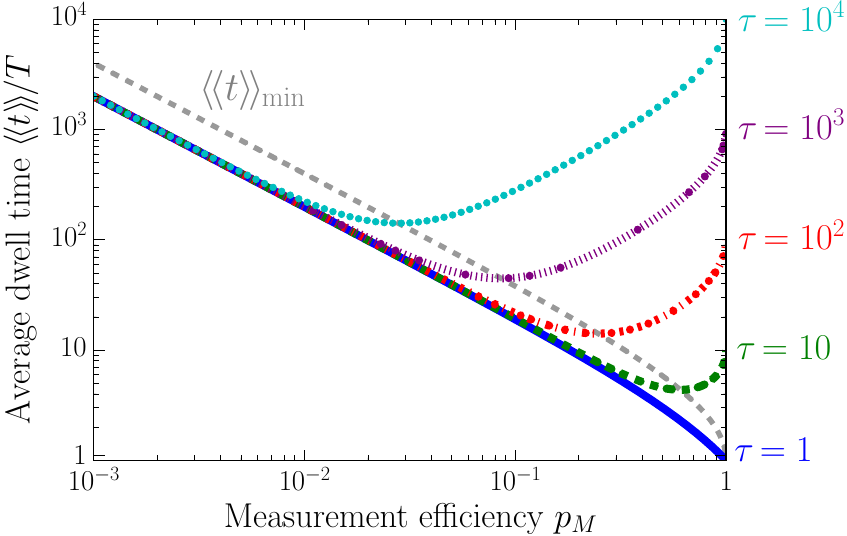}
 \caption{Plot of the formula~\eqref{eq:dwell_time_result} for the double-averaged dwell time $\dblavg{t}$ as a function of the measurement efficiency $p_M$, for different values of the quantity $\tau$ defined in Eq.~\eqref{eq:decay_time_pm=1}. The grey dashed line shows the minimal average dwell time, obtained by substituting Eq.~\eqref{eq:pm_opt} for different $\tau$. Due to the quantum Zeno effect, $\tau = \dblavg{t}|_{p_M=1}$ can be much larger than $\dblavg{t}|_\text{min}$.}
  \label{fig:decay_time}
 \end{figure}

A counterintuitive way to speed up the measurement process is to \emph{decrease} the measurement efficiency $p_M$.
As the first term in Eq.~\eqref{eq:dwell_time_result} shows, for
$p_M\approx 1$, decreasing $p_M$ decreases the dwell time, in
close analogy with the quantum Zeno Effect.
The price to pay for weak measurements is the second term of
Eq.~\eqref{eq:dwell_time_result}, which diverges in the limit
$p_M\to 0$ as $\propto1/p_M$. 
Hence, there is an optimal value of $p_M$ that minimizes $\dblavg{t}$, given by
\begin{equation}\label{eq:pm_opt}
p_M^* =2 \,\frac{
  \tau\sqrt{2\tau-1} - (2\tau-1)}{(\tau-1)^2}.
\end{equation}
For $\tau\gg 1$, this optimal choice of $p_M = p_M^\ast \approx \sqrt{8/\tau}$ reduces the time needed to
perform the measurement from $T \tau$ to $\dblavg{t}|_\text{min}
\approx T \sqrt{2\tau}$, which is a speed-up by a factor of
$\mathcal{O}(\sqrt{\tau})$. These results are illustrated in 
Fig.~\ref{fig:decay_time}. 

Translating our results for the dwell time to the
nondriven case of Eq.~\eqref{eq:nondriven_def}, where $\gamma$ is the
loss rate, we find 
\begin{align}
\label{eq:dwell_time_nondriven}
\dblavg{t} &= \frac{\gamma \left.\mathfrak{t}\right.^2}{4} + \frac{2}{\gamma}, &
\left.\mathfrak{t}\right.^2 &\equiv \frac{1}{N} \int_0^\infty \frac{\rho(E)}{E^2} \mathrm{d}E.
\end{align}
The quantum Zeno-like effect applies here too. The
integral can diverge due to dark and almost-dark states at
$E\approx 0$.
The optimal choice of the decay rate is
$\gamma=\sqrt{8}/\mathfrak{t}$, when the two terms of Eq.~\eqref{eq:dwell_time_nondriven} are equal.
In this case $\dblavg{t}=\sqrt{2}\,\mathfrak{t}$.
For the translationally invariant lossy SSH model of
Ref.~\onlinecite{rudner2009topological}, we find
$\mathfrak{t}=|v^2-{v'}^2|^{-1/2}$, whereby for $v\approx v'$
we obtain $ \dblavg{t} \propto 1/\abs{v-v'}$ as in 
Ref.~\cite{rudner2009topological}.
We note that also the usual quantum Zeno effect shows up here: in the
limit $\gamma\to\infty$ the measurement process slows down instead of
speeding up, with the dwell time diverging as $\dblavg{t} \propto
\gamma$.

\section{Experimental proposal} Our results could
be tested in a variety of experimental settings.
One possibility consists in using photonic devices, modifying the experimental scheme of Refs.~\cite{zeuner2015observation,Bleckmann:2016} based on evanescently coupled waveguides, to include periodic dynamics.
Instead of continuous losses from the fast modulation of the waveguides \cite{zeuner2015observation}, discrete loss operations after each driving cycle can be implemented by discrete beam-splitter elements, e.g., by laser writing in glass substrates \cite{sciarrino_twoparticle}.
These coherently redirect a portion $p_M$ of light from one of the two sublattices into auxiliary spatial modes, which constitute the ancillary sites for the weak measurements of position.
Recording their intensity distribution at the ends of the waveguide array allows the average displacement to be computed, as defined in Eq.~\eqref{eq:single_avgs1}.

An alternative, promising experimental scheme to test our results relies on a discrete-time quantum walk \textemdash a quantum particle with internal states moving in discrete steps on a one-dimensional lattice.
These are ideal systems to study periodically driven quantum systems where the effective Hamiltonian cannot be simply obtained perturbatively from the time-averaged Hamiltonian \cite{Goldman:2014};
moreover, they have the advantage of realizing chiral symmetry exactly, a condition which is hard to guarantee in coupled waveguide arrays \cite{Bleckmann:2016}.
In the past years, they have been realized using trapped ions~\cite{roos_ions}, cold atoms in optical lattices~\cite{karski2009quantum,alberti_nonclassicality}, pulses of light~\cite{schreiber2010photons,peruzzo_science_2010,
 kitagawa2012observation,sciarrino_twoparticle,cardano2016detection}, and most recently using superconducting devices~\cite{flurin2016observing,ramasesh2016direct}.
In particular, ultracold atoms trapped in polarization-synthesized optical lattices \cite{groh_robustness} are ideal candidate to test our results.
We have recently demonstrated that negative measurements of the atom's position can be realized using long spin-selective shift operations  \cite{alberti_nonclassicality,Robens:2016,BasisChangeNote}.
The spatial distribution of the removed atoms can be recorded via fluorescence imaging \cite{Alberti:2016} after the last step.

To provide a numerical example, we choose the split-step quantum-walk protocol \cite{kitagawa2010topological}, which is the simplest one to possess a rich variety of topological phases.
We consider $2N=2$ internal states, denoted by $\ket{\uparrow}$ ($c=1$) and $\ket{\downarrow}$ ($c=0$), which we refer to as spin.
The operator describing the unitary evolution of a single step of the walk is defined as
\begin{align}
\label{eq:U_splitstep_def}
\UU(\theta_1,\theta_2) &= \hat{R}(\theta_1/2) \hat{S}_{-} \hat{R}(\theta_2) \hat{S}_{+} \hat{R}(\theta_1/2),
\end{align}
where $\hat{R}(\theta)$ rotates the spin around the $y$-axis by an angle $\theta(x)$, depending in general on site $x$, and $S_\uparrow$ ($S_\downarrow$) shifts the particle by $+1$ ($-1$) site if the internal state is $\ket{\uparrow}$ ($\ket{\downarrow}$), leaving it unaffected otherwise.
The topological invariants depend on the rotation angles $\theta_1$ and $\theta_2$, and are well known for both translationally invariant~\cite{kitagawa2010topological,asboth2013bulk} and spatially disordered angles~\cite{tarasinski2014scattering,rakovszky2015localization}.
The chiral-symmetry operator is $\GG = \hat{1} \otimes \sigma_x$, thus the two sublattices correspond to the two internal states \encapsulateMath{$\ket{\kern-3pt\begin{array}{c}\rightarrow\\[-2.75mm]\leftarrow \end{array}\kern-3pt}=(\ket{\uparrow}\pm\ket{\downarrow})/\sqrt{2}$}.
After each unitary step, we remove the particle in the state $\ket{\leftarrow}$, and records its position (assuming $p_M=1$ for simplicity).

\begin{figure}[!ht]
   \centering
  		\includegraphics[width=0.475\textwidth]{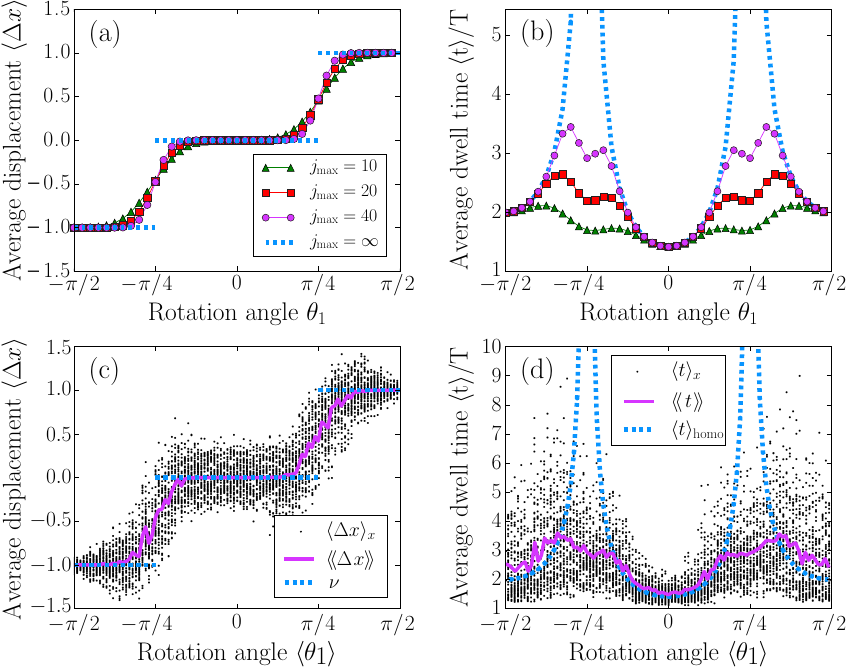}

 \caption{Average displacement and dwell time for the split-step quantum walk with strong measurement ($p_M = 1$) as the coin angle $\theta_1$ is tuned ($\theta_2$ centered at $\pi/4$, $L=50$ sites).
 Step changes of the average displacement indicate topological phase transitions.
 (a)-(b) A homogeneous split-step walk with the time evolution terminated after $j_\text{max}$ steps.
 (c)-(d) Split-step walk with disordered rotation angles uniformly distributed in intervals of width $\pi/5$ (see text), with total number of steps $j_\text{max} = 40$. Black dots correspond to different initial sites, and the red curve represents their average.
 }
  \label{fig:splitstep}
 \end{figure}

The average displacement and the average dwell time are shown in Fig.~\ref{fig:splitstep} for numerical examples, both with and without spatial disorder. 
Here the time evolution is terminated after $j_\text{max}$ steps.
In the translationally invariant case, for parameters far from topological phase transitions, $j_\text{max} = 10$ steps are sufficient to observe the quantized displacement predicted by Eq.~\eqref{eq:mainstatement}.
Close to a phase transition, the average dwell time becomes large, as shown by Eq.~\eqref{eq:decay_time_pm=1}, and the quantization of the displacement breaks down due to finite $j_\text{max}$.
In the disordered case, shown in Fig.~\ref{fig:splitstep} (c) and (d), we use rotation angles chosen uniformly from the intervals $\theta_{1,2}\in[\expect{\theta_{1,2}}-\pi/10,\expect{\theta_{1,2}}+\pi/10]$.
We fix $\expect{\theta_2} = \pi/4$ and tune $\expect{\theta_1}$.
The displacements corresponding to different initial states are no longer quantized, but by averaging over them we recover the quantized winding number, for time evolution terminated after $t_\text{max} = 40$ steps.

\section{Discussion and conclusions} We proved that losses can be used to
detect bulk topological invariants in chiral symmetric one-dimensional
lattices, with any number of internal states, disordered or
translationally invariant, periodically driven or static.
This is a powerful generalization of some of the results of Rudner and
Levitov on the SSH model; as in their case, we expect that it should
even be possible to relax the requirement of chiral
symmetry and allow for certain types of decoherence~\cite{rudner2009topological,rudner2016survival}.
This approach should also be useful to obtain (weak) topological
invariants of chiral symmetric systems in two dimensions and above.
Exploring the relations between our results and the inspiring recent
work by Cardano et
al~\cite{cardano2016statistical,cardano2016detection} on the
periodically driven SSH model would be an interesting topic for
future research.

\begin{acknowledgments}
We acknowledge useful discussions with Mark Rudner. 
J.K.A. acknowledges support from the Hungarian Scientific
Research Fund (OTKA) under Contract No. NN109651, and from the Janos
Bolyai Scholarship of the Hungarian Academy of Sciences.
AA also acknowledges financial support from the the ERC grant DQSIM
and from the Deutsche Forschungsgemeinschaft SFB TR/185 OSCAR.
\end{acknowledgments}

\bibliographystyle{apsrev4-1}

\bibliography{walkbib}{}

\begin{thebibliography}{52}
\bibitem [{Bas()}]{BasisChangeNote}%
  \BibitemOpen
  \bibinfo {note} {A basis change by $\pi/2$ microwave pulses is required to
  selectively remove $\ket{\leftarrow}$.}\BibitemShut {Stop}%
\end{thebibliography}

\end{document}